\documentclass[11pt]{article}
\usepackage{moriond,epsfig}

\def\be{\begin{equation}}
\def\ee{\end{equation}}
\def\bea{\begin{eqnarray}}
\def\eea{\end{eqnarray}}

\begin{document}
\vspace*{4cm}
\title{Looking for dark matter annihilations in dwarf galaxies \footnote{Contribution to the proceedings of the 2004 Moriond workshop ``Exploring the Universe''. Based on~\cite{ferrer1}.}}

\author{F. Ferrer}

\address{Astrophysics, University of Oxford,\\ Keble Road,
Oxford OX1 3RH, UK}

\maketitle\abstracts{We calculate the flux of high energy
$\gamma$-rays from annihilation of neutralino dark matter in the
centre of the Milky Way and the three nearest dwarf spheroidals
(Sagittarius, Draco and Canis Major), using {\em realistic} models of the
dark matter distribution.}

\section{Introduction}
 The term Dark Matter~(see e.g.~\cite{Bertone:2004pz} for a review) denotes {\it any form of matter whose presence is inferred solely from its gravitational effects}. The first evidence for its existence was found by Zwicky who noticed in 1933 that the visible galaxies in the Coma cluster could account for only about a tenth of the total mass binding the cluster. In 1974, Ostriker et al. and Einasto et al. pointed out the need for large amounts of dark matter around isolated galaxies. Moreover, current models of structure formation generically require the presence of cold, non-relativistic, dark matter (CDM) for the primordial fluctuations to grow into the galaxies that we see nowadays. The nature of this CDM remains largely unknown although the success of Big Bang Nucleosynthesis implies that most of it is non-baryonic.

The foremost candidate for CDM composing
galactic haloes is the lightest supersymmetric particle, which in some
popular models of softly broken supersymmetry (e.g. mSUGRA) turns out
to be the neutralino. If so, then neutralino pair annihilation may
lead to observable consequences, in particular the emission of high
energy $\gamma$-radiation. The possibility that such $\gamma$-rays may
be identified by forthcoming atmospheric Cerenkov telescopes (ACT)
such as VERITAS or by satellite-borne detectors like GLAST has excited
considerable recent interest.

\section{The local distribution of Dark Matter}

It is clearly of importance to identify the best places to search for
such an annihilation signal. Inspired by the highly cusped models
based on numerical simulations of dark halo formation \cite{navarro},
a number of investigators have suggested that the centre of the Milky
Way may be the optimum target. If the dark matter density is cusped as
$1/r$ at small radii, then the $\gamma$-ray flux would be detectable
for typical neutralino properties in the minimal supersymmetric
extension of the Standard Model~\cite{Bergstrom98}. 

Awkwardly, there is a substantial body of astrophysical evidence that
the halo of the Milky Way is not cusped at all.
First, the microlensing optical depth towards the Galactic Center is
very high. Particle dark matter does not cause microlensing, whereas
faint stars and brown dwarfs do. The total amount of all matter within
the Solar circle is constrained by the rotation curve, so this tells
us that lines of sight towards the Galactic Center are not dominated
by particle dark matter. Haloes as strongly cusped as $1/r$,
normalised to the local dark matter density as inferred from the
stellar kinematics in the solar neighbourhood, are ruled out by the
high microlensing optical depth \cite{be}. Second, the pattern speed
of the Galactic bar is known to be fast from hydrodynamical modelling
of the motions of neutral and ionised gas. If dark matter dominates
the central regions of the Milky Way, then dynamical friction will
strongly couple the dark matter to the Galactic bar and cause it to
decelerate on a few bar rotation timescales \cite{debs}. It is now
largely accepted by astronomers that bright galaxies like the Milky
Way do not have cusped dark haloes today. For the three nearest dSphs
-- Draco, Sagittarius and Canis Major -- there is no direct evidence
either for or against central cusps in their dark matter distribution.

Dwarf spheroidals (dSphs) are amongst the most extreme dark matter
dominated environments. The mass-to-light ratio of Draco is $\sim 250$
in Solar units, while that of the Sagittarius is $\sim 100$. The
recently discovered possible dSph in Canis Major seems similar to the
Sagittarius in structural properties and dark matter content.  Given
the seeming absence of dark matter in globular clusters, dSphs are
also the smallest systems dominated by dark matter.

We develop two sets of models of dSphs~\cite{ferrer1}: cored spherical
power-law models and cusped haloes favored by numerical simulations.

The shape of the profile is determined by fitting~\cite{ferrer1} to
observational data on the Draco dSph using the Jeans equation
\cite{bt}. For a spherical galaxy, the enclosed mass $M(r)$ is related
to observables via
\begin{equation}
M(r) = 
 - \frac{r\langle v_r \rangle^2}{G}
 \left(\frac{{\rm d}\log\nu}{{\rm d}\log r} 
 + \frac{{\rm d}\log\langle v_r^2 \rangle}{{\rm d}\log r} 
 + 2\beta\right).
\label{eq:fitgen}
\end{equation}
Here, $\nu$ is the luminosity density, $\langle v_r^2 \rangle$ is the
radial velocity dispersion of the stars and $\beta$ is the anisotropy
of the stellar motions. 

To determine the extent of the dark matter halo of the dSphs, the
tidal radius must be estimated.  The approximate method used
conventionally is derived from the Roche criterion. The tidal radius
is found by requiring that the average mass in the dSph is equal to
the average interior mass in the Milky Way halo.

\section{The Gamma-ray flux}

The $\gamma$-ray flux from neutralino annihilation is given by
\cite{Bergstrom98}%
\begin{equation}
\Phi_\gamma(\psi) = \frac{N_\gamma\langle\sigma v\rangle}{4\pi m_\chi^2} 
 \times \frac{1}{\Delta\Omega} \int_{\Delta\Omega} {\rm d}\Omega
 \int_{\rm los} \rho^2 [r(s)]\ {\rm d}s,
\label{eq:integrand}
\end{equation}
where $m_\chi$ and $\langle \sigma v \rangle$ are the neutralino mass
and its self-annihilation cross-section, $\rho$ is the density of the
dSph and the integration is performed along the line-of-sight to the
target and averaged over the solid angle $\Delta\Omega$. $N_\gamma$ is
the number of photons produced in each annihilation process.

We focus on minimal supergravity (mSUGRA) models with universal
gaugino and scalar masses and trilinear terms at the unification
scale. We use the computer programme SoftSusy~\cite{softsusy} to scan
the supersymmetric parameter space. The output at the electroweak
scale is fed into the programme DarkSusy~\cite{ds} which computes the
relic density and products of the neutralino annihilations. It also
checks that a given model is not ruled out by present accelerator
experiments.  A feasible model is one which is permitted by
accelerator limits and which predicts a relic density in the range
$0.005 < \Omega_{\rm CDM} h^2 < 0.2$.

Typical reference sizes for the solid angle are
$\Delta \Omega=10^{-5}$ sr for ACTs and GLAST and $\Delta
\Omega=10^{-3}$ sr for EGRET. EGRET and GLAST are satellite
detectors with low energy thresholds ($\approx 100$ MeV), high energy
resolution ($\approx 15\%$) but only moderate angular precision. The
others are ACTs with higher thresholds ($\approx$ 100 GeV) but better
angular resolution.  

The minimum detectable flux $\Phi_\gamma$ is determined using the
prescription that, for an exposure of $t$ seconds made with an
instrument of effective area $A_{\rm eff}$ and angular acceptance
$\Delta \Omega$, the significance of the detection must exceed $5
\sigma$, i.e. $\frac{\Phi_\gamma \sqrt{\Delta\Omega A_{\rm eff}
t}}{\sqrt{\Phi_\gamma+\Phi_{\rm bg}}} \ge 5$.  Here, $\Phi_\gamma$
denotes the neutralino annihilation flux, while $\Phi_{\rm bg}$ is the
background flux. There are three sources of background for the signal
under consideration: hadronic, cosmic-ray electrons and diffuse
$\gamma$-rays from astrophysical processes.  The last is negligible
for ACTs, but is the only one present for satellite experiments like
GLAST or EGRET.

\begin{figure}
\begin{center}
\begin{tabular}{c@{\hspace{2cm}}c}
\epsfig{figure=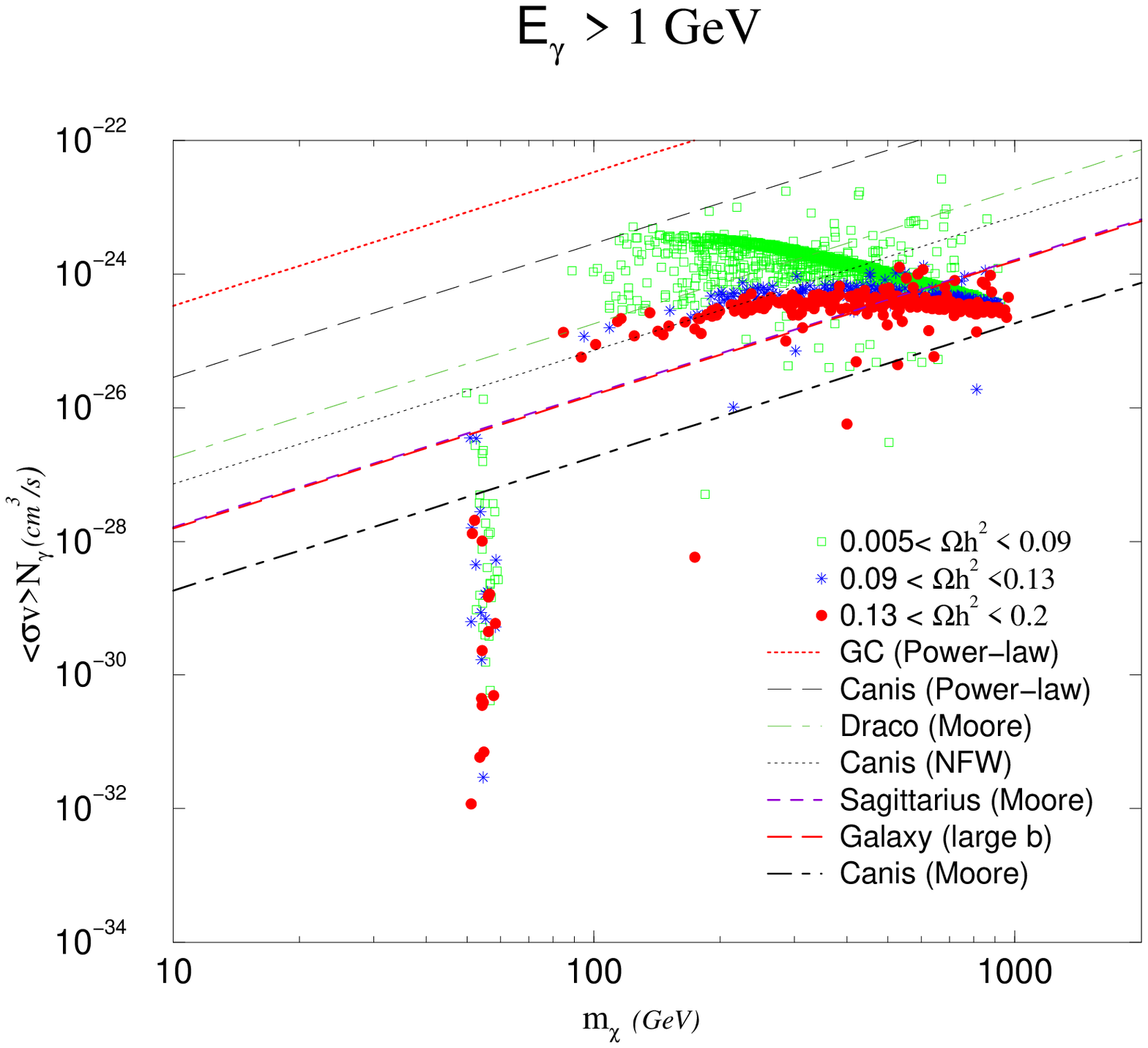,height=2.in}&
\epsfig{figure=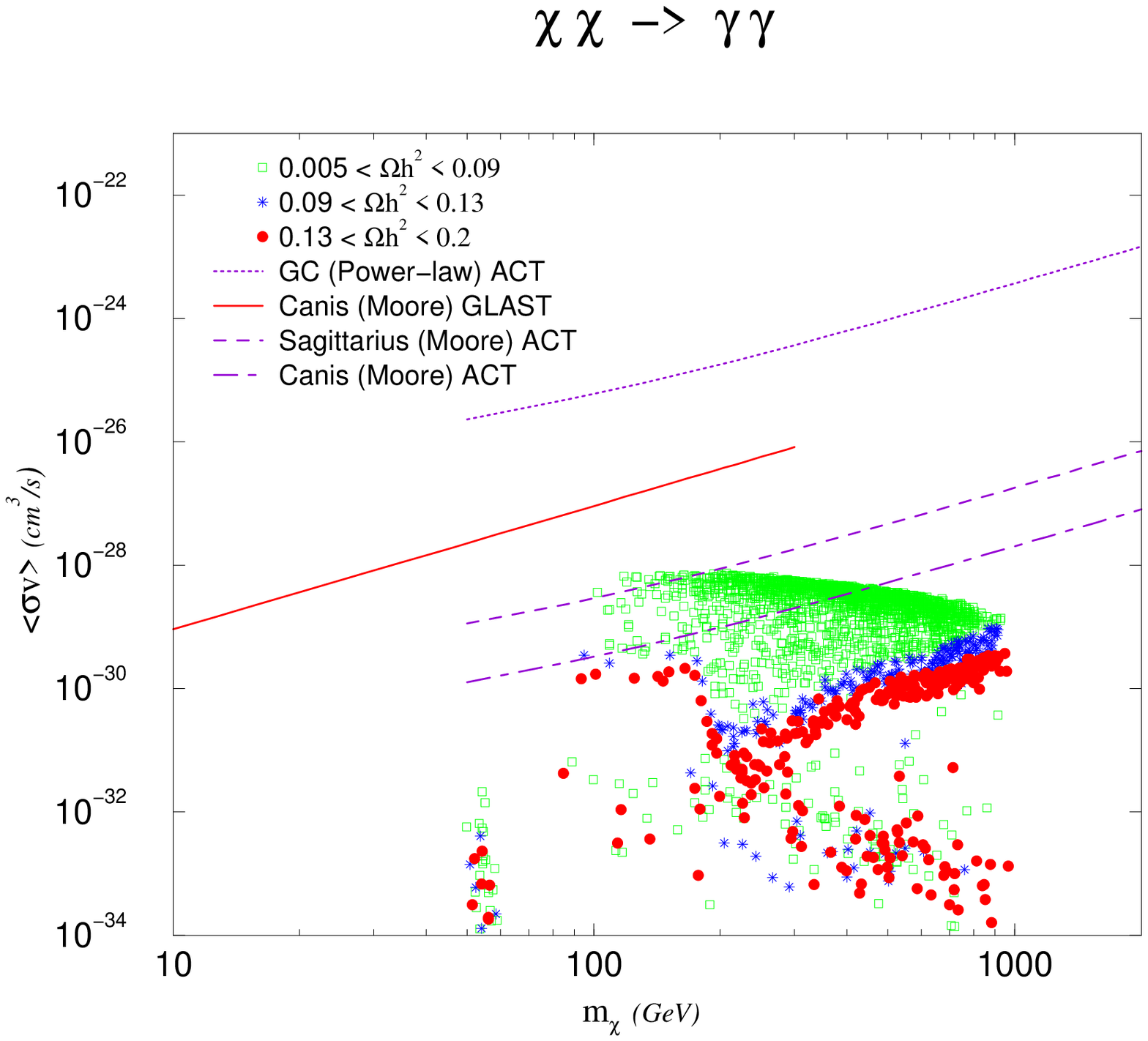,height=2.in}
\end{tabular}
\end{center}
\caption{Exclusion limits for continuum $\gamma$-ray emission above 1
GeV (left) and for the discrete line $\chi \chi \rightarrow \gamma
\gamma$ (right).
\label{fig:excl}}
\end{figure}

Fig.~\ref{fig:excl} shows the parts of the supersymmetric parameter
space that can be probed through the detection of a $\gamma$-ray
signal from neutralino annihilations. We show the region to
which GLAST and a generic second generation ACT will be sensitive.

The discrete annihilation line is very unlikely
to be observed, even with the next generation instruments. It is just
about detectable for the most promising targets under the most
optimistic assumptions -- the Sagittarius or the Canis Major dSph
galaxies assuming a Moore profile and using next generation ACTs.
Other possible models (such as NFW or cored profiles) and targets
(such as the Galactic Center) are much less propitious still.

The continuum emission comes from hadronization and subsequent pion
decay. The Draco, Sagittarius and Canis Major dSphs may yield
interesting constraints -- but only if their dark halo profiles are
strongly cusped. Unlike the case of the Milky Way, cusped profiles are
still possible for the dSphs. For $E_\gamma > 1$ GeV, only curves for
GLAST are drawn, as ACTs are insensitive at such low energies.

Also shown is a line corresponding to the Milky Way observed at medium
latitudes with the wide field of view of GLAST, as first suggested by
Stoehr et al. \cite{stoehr}. Here, the Galaxy has
been modelled with an isothermal power-law model, as opposed to the
cusped models preferred by Stoehr {\it et al.} This is a promising target,
as {\em irrespective of whether the Galaxy is cusped or cored}, there
are always useful constraints on the supersymmetric
parameters. Unfortunately, this attractive option is only available to
GLAST and not for ACTs.

\section{Conclusions}

If the dark matter present in the Universe is composed by the lightest
supersymmetric particle, then this could manifest itself via
$\gamma$-ray emission from pair annihilations. There have been a
number of recent calculations predicting that the neutralino
annihilation flux from the inner Galaxy will be detectable with
forthcoming satellites like GLAST and with second generation
atmospheric Cerenkov telescopes (ACTs). These calculations assume that
the cusped Navarro-Frenk-White (NFW) models for the Milky Way halo
hold good. This assumption is in contradiction with a substantial body
of astrophysical evidence about the inner Galaxy \cite{be,debs}.

The high mass-to-light ratios of the Local Group dwarf spheroidals
(dSphs) makes them attractive targets.  Cusped profiles like NFW are
not presently ruled out for dSphs like Sagittarius or Draco. The
detection of monochromatic lines is still extremely difficult, but the
GLAST satellite may detect excess continuum $\gamma$ ray emission. 

\section*{Acknowledgments}
This work was done in collaboration with N.~W.~Evans and S.~Sarkar and supported by the Leverhulme trust.


\section*{References}
\begin{thebibliography}{99}
\bibitem{ferrer1}
N.~W.~Evans, F.~Ferrer and S.~Sarkar,
Phys.\ Rev.\ {\bf D69}, 123501 (2004)
[arXiv:astro-ph/0311145].

\bibitem{Bertone:2004pz}
G.~Bertone, D.~Hooper and J.~Silk,
arXiv:hep-ph/0404175.

\bibitem{navarro}
J.~F.~Navarro, C.~S.~Frenk and S.~D.~White,
Astroph.\ J.\ {\bf 490}, 493 (1997)
[arXiv:astro-ph/9611107];
B.~Moore, F.~Governato, T.~Quinn, J.~Stadel and G.~Lake, 
Astrophys.\ J.\  {\bf 499}, L5 (1998)
[arXiv:astro-ph/9709051].

\bibitem{Bergstrom98}
L.~Bergstr\"om, P.~Ullio and J.~H.~Buckley,
Astropart.\ Phys.\ {\bf 9}, 137 (1998)
[arXiv:astro-ph/9712318].

\bibitem{be}
J.~J.~Binney and N.~W.~Evans,
Mon.\ Not.\ R.\ Astron.\ Soc.\ {\bf 327}, L27 (2001)
[arXiv:astro-ph/0108505].

\bibitem{debs}
V.~Debattista and J.~A.~Sellwood,
Astrophys.\ J.\ {\bf 493}, L5 (1998)
[arXiv:astro-ph/9710039].

\bibitem{bt}
J.~Binney, S.~Tremaine, 
``Galactic Dynamics'', 
Princeton University Press, 1987 (section 4.2).




\bibitem{softsusy}
B.~C.~Allanach, 
Comput.\ Phys.\ Commun.\ {\bf 143}, 305 (2002)
[arXiv:hep-ph/0104145]. 
{\tt http://allanach.home.cern.ch/allanach/softsusy.html}.

\bibitem{ds}
P.~Gondolo, J.~Edsj\"o, L.~Bergstrom, P.~Ullio and E.~A.~Baltz,
{\tt http://www.physto.se/$\sim$edsjo/darksusy/}


\bibitem{stoehr}
F.~Stoehr, S.~D.~White, V.~Springel, G.~Tormen and N.~Yoshida,
Mon.\ Not.\ Roy.\ Astron.\ Soc.\  {\bf 345}, 1313 (2003)
[arXiv:astro-ph/0307026].


\end{thebibliography}
\end{document}